\documentclass[twocolumn,superscriptaddress,showpacs,pra]{revtex4}
\usepackage{epsfig}
\usepackage{delarray}
\usepackage{amsmath, amssymb}
\usepackage{bm}
\newcommand{\ket}[1]{| \, #1 \rangle}

\usepackage{epstopdf}

\begin{document}
\title{Fusing multiple W states simultaneously with a Fredkin gate}

\author{Fatih Ozaydin}
\email{MansurSah@gmail.com}
\affiliation{Department of Information Technologies, Isik University, Istanbul, Turkey}

\author{Sinan Bugu}
\affiliation{Department of Computer Engineering, Istanbul University, Istanbul, Turkey}

\author{Can Yesilyurt}
\affiliation{Department of Computer Engineering, Okan University, Istanbul, Turkey}

\author{Azmi Ali Altintas}
\affiliation{Department of Electrical Engineering, Okan University, Istanbul, Turkey}

\author{Mark Tame}
\email{markstame@gmail.com}
\affiliation{School of Chemistry and Physics, University of KwaZulu-Natal, Durban 4001, South Africa}

\author{\c{S}ahin Kaya \"Ozdemir}
\email{ozdemir@ese.wustl.edu}
\affiliation{Department\,of\,Electrical\,and\,Systems\,Engineering, Washington University, St. Louis, MO 63130, USA}

\date{\today}
\begin{abstract}
We propose an optical scheme to prepare large-scale entangled networks of W states. The scheme works by simultaneously fusing three polarization-encoded W states of arbitrary size via accessing only one qubit of each W state. It is composed of a Fredkin gate (controlled-swap gate), two fusion gates [as proposed in {\it New J. Phys. 13, 103003 (2011)}] and an H-polarized ancilla photon. Starting with three $n$-qubit W states, the scheme prepares a new W state with $3(n-1)$-qubits after postselection if both fusion gates operate successfully, {\it {\it i.e.}} a four-fold coincidence at the detectors. The proposed scheme reduces the cost of creating arbitrarily large W states considerably when compared to previously reported schemes.

\end{abstract}
\maketitle


\section{Introduction}

Entanglement between two or more particles has attracted a great deal of attention since the seminal EPR paper~\cite{EPR1935}. Researchers have studied entanglement by exploring its multitude of fundamental quantum features as well as investigating it from an information theoretical perspective~\cite{Horodecki2009}. The creation, manipulation and characterization of bipartite entanglement has now been achieved in a wide range of physical settings~\cite{Gisin2002,Giovannetti2006,Kimble2008,Horodecki2009,Ladd2010}. Here, bipartite entangled states have been put into practical use for studying fundamental concepts in quantum mechanics and for realizing tasks which cannot be achieved in classical systems, such as teleportation~\cite{Bennett1993} and quantum key distribution~\cite{Gisin2002}. Despite the remarkable theoretical and experimental progress that has been made in the study of bipartite entangled states, more general multipartite entangled states are yet to be fully explored. This is mainly due to the added complexities involved in their mathematical formulation and characterization, as well as the rapidly increasing number of resources needed to prepare large-scale entangled states in experiments.

A striking difference between bipartite and multipartite entanglement is the presence of inequivalent classes of entanglement in the latter. While any bipartite state can be obtained by local operations and classical communication (LOCC) from a bipartite entangled state, the same is not possible for multipartite entanglement. A multipartite entangled state belonging to one of the inequivalent classes cannot be converted to states in other classes with LOCC~\cite{Dur2000,Acin2001,Verstraete2002}. Well known examples of such classes are GHZ~\cite{GreenbergerarXiv,Greenberger1989}, W~\cite{Dur2001}, Dicke~\cite{Dicke1954} and cluster states~\cite{Briegel2001}. Studies have shown that each of these inequivalent classes plays an important role in uncovering the fundamentals of entanglement in quantum information science and in realizing a range of quantum information processing tasks. For example, cluster states have been shown to be a universal resource for measurement-based quantum computing, GHZ states are known to be required for reaching consensus in quantum networks under specific constraints, and W states are required for leadership election in anonymous quantum networks~\cite{Hondt2006} and for secure quantum communication~\cite{Joo2002,Wang2007,Cao2006,Liu2011}. The efficient preparation of large-scale multipartite entangled states of certain classes is therefore a crucial step in quantum information processing. However, as their structure becomes sophisticated and the number of qubits increases, the preparation and characterization of multipartite entangled states becomes challenging.

In recent work the concept of creating cluster states with a large number of qubits by fusing cluster states with a smaller number of qubits was proposed~\cite{NJPRef13}. This work was followed by an increasing number of theoretical proposals for constructing {\it quantum fusion gates} to prepare large entangled states and the analysis of their optimality under various fusion scenarios~\cite{fusionopt}. Efficient preparation of cluster states and GHZ states via the fusion process have been well-investigated theoretically~\cite{NJPRef13,fusionopt,NJPRef14} and demonstrated in experiments~\cite{Zhang2006,Bell2012}. Extending the concept of state fusion to other classes of entanglement, such as to Dicke and W states is still in its early stage. In the study of the fusion process for different classes of entanglement the key constraint is the number of accessible qubits, {\it {\it i.e.}} how many qubits from each of the states are to be sent to the fusion gate. It is generally desired that the fusion operation succeeds by accessing only one qubit from each state as this reduces the complexity of the fusion operation, making it more experimentally feasible. 
Here, studies have shown that two W states, GHZ states or cluster states can be fused together to form a larger state with the same entanglement structure as the input states by accessing only one qubit from each of the initial states and these initial states can also be expanded by adding one or two qubits at a time by locally accessing only one of their qubits~\cite{Sahin-NJP-Oct,NJPRef13,Lu2007NatPhys,NJPRef31,NJPRef32,Ikuta2011}. However, for Dicke states this is not the case~\cite{Kobayashi2013} and there has not yet been a proposal for such a fusion gate for these states.

In general, the efficiency of the fusion or expansion process and the cost of preparing an entangled state with a desired number of qubits depend on the structure of the entangled states, as well as the complexity of the inner workings of the fusion gate. In linear optical systems, fusion or expansion gates are usually non-deterministic, that is, the process prepares the desired state only probabilistically. Fortunately, in most fusion gate proposals, there is a case where even if the fusion gate fails to prepare the desired state, the failure does not destroy the initial entangled states completely. This case is termed the `recyclable' case, because although each qubit from the initial states entering the fusion gates is destroyed, the remaining ($N-1$) qubits of each of the states still keep their entanglement structure intact so that a new round of fusion can be performed on them. Availability of such recyclable cases increases the efficiency of the process and reduces the cost of preparing the desired state.

In this work we focus on fusing $n$-qubit W states, defined as the summation of all possible $n$-qubit states with one excitation in spin up, {\it {\it i.e.}} $\ket{W_n}=\frac{1}{\sqrt{n}}\sum_{perm}\ket{00\dots 01}$, where $\{ \ket{0}, \ket{1} \}$ represents the single-qubit computational basis. In particular we consider a photonic setting as it represents a natural choice for the distribution of such entangled states over communication networks~\cite{Gisin2002,Kimble2008}. In this setting, we use the polarization degree of freedom of the photons to embody the qubits, where $\{ \ket{1_H},\ket{1_V} \} \leftrightarrow \{ \ket{0}, \ket{1} \}$. Already quite a few theoretical proposals and experimental implementations have been carried out for expanding and fusing photonic W states: an optical gate has been proposed for expanding a given W state by two qubits~\cite{NJPRef31}, and a simpler gate which expands a W state by one qubit was proposed and shown to be realizable with current photonic technology~\cite{NJPRef32}. The maximum success probability of the expansion has been derived~\cite{Ikuta2011} and the expansion of a W state has been experimentally realized, generating W states of three and four photons~\cite{NJPRef33}. Furthermore, two EPR pairs have been experimentally transformed into a W state of three photons~\cite{NJPRef26}. In all these methods, only one qubit of the initial state is accessed. Therefore the trade off between the improvement of the process and accessing more qubits (requiring more complex gates) remains an interesting open problem.

Here, we propose a scheme to fuse W states but with a much simpler setup compared to earlier work~\cite{Yesilyurt2013A} and with a better resource cost for any target size when compared to the schemes in Refs.~\cite{Yesilyurt2013A} and~\cite{Sahin-NJP-Oct}. This new fusion scheme consists of a Fredkin gate and two fusion gates, and uses a single ancilla photon. It fuses three W states of arbitrary size with the help of the ancilla photon, and requires access to only one qubit of each of the W states. The same scheme can also be used to fuse Bell states to prepare W states with larger numbers of qubits. Our proposed scheme reduces the cost of creating arbitrarily large W states considerably when compared to previously reported schemes.

The paper is organised as follows. In Section II we provide background theory of previous W state preparation schemes using fusion operations and the drawbacks of these schemes. In Section III we then introduce our proposed fusion scheme, outlining in detail how it works. In Section IV we analyse the efficiency of our scheme compared to earlier works and calculate the resource cost of preparing large W states. Finally, in Section V we summarize our main results and provide an outlook for future work.


\section{W state preparation via fusion}

In previous studies the main problem of preparing large-scale W states was that the resource cost increased exponentially with respect to the target state size, due to the low probability of success of the setups proposed. A simple optical fusion gate has recently been introduced (though not experimentally demonstrated yet) which can fuse two arbitrary size W states and create a W state with a resource requirement increasing sub-exponentially with respect to the target size~\cite{Sahin-NJP-Oct}. This \textit{``fusion gate''} for W states consists of a polarization beam splitter (PBS), a half-wave plate (HWP) and two photon detectors. The fusion is performed as follows: Alice and Bob have W states of sizes $n$ and $m$, {\it i.e.} $|W_n\rangle_{A}$ and $|W_m\rangle_{B}$ and they wish to fuse their states. Each sends one photon to the fusion gate and there appear four cases, $\{\ket{1_H}\ket{1_H}~\to~\ket{0}\ket{1_H1_V}$, $\ket{1_H}\ket{1_V}~\to~\ket{1_H}\ket{1_H}$, $\ket{1_V}\ket{1_H} \to \ket{1_V}\ket{1_V}$, $\ket{1_V}\ket{1_V} \to \ket{1_H1_V}\ket{0} \}$ with the associated probabilities $ \{ {(n-1)(m-1) \over nm}, {n-1 \over nm}, {m-1 \over nm}, {1 \over nm} \}$, respectively. The photons in the output modes are measured in the basis $\{ \ket{1_+}, \ket{1_-}\}$, where $\ket{1_\pm}=(\ket{1_H}\pm\ket{1_V})\sqrt{2}$ so that a coincidence detection can only have come from the case when both input photons have orthogonal polarization, with the detectors unable to distinguish between the two cases. The fusion gate is postselective as the detection outcomes are non-deterministic. Thus, the last case is a `failure' which destroys the entanglement in the remaining qubits regardless of the number of qubits in the initial W states. The indistinguishable second and third cases are successful outcomes resulting in a genuine W state of size $n+m-2$. In the first case, which is termed as a recyclable outcome, each party loses one photon but is left with a W state of a smaller size. Therefore they may try to repeat the process until the size of any party decreases to two, which is a \textit{W-type} Bell state ($\ket{W_2}=\ket{\psi^+}$). Note that no growth can be achieved if $n$ or $m$ equals to $2$. Therefore to create $|W_3\rangle$ states, one would also need the gate of~\cite{NJPRef26} for example, not only for starting the process but also for any strategy including recycling, whenever the size of any party decreases to $2$. In order to compare various strategies, a resource cost of preparing a W state of size $k$, $R[W_k]$ is defined
\begin{equation}
R[W_{n+m-2}] = { R[W_n] + R[W_m] \over P_s(W_n,W_m)}
\end{equation} where $P_s(W_n,W_m)$ is the probability of success of fusing the states $W_n$ and $W_m$. The cost of the primary resource, is assumed to have unit cost, {\it i.e.} $R[W_3]=1$.

\begin{figure}[t]
  \centering
      \includegraphics[width=0.35\textwidth]{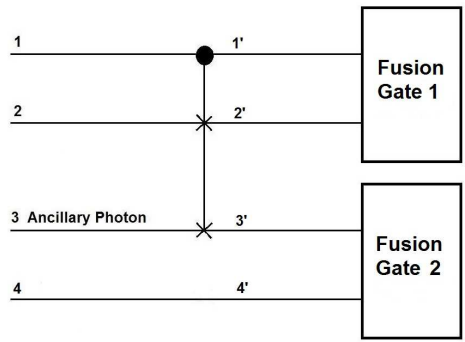}
  \caption{\label{fig:AmeleyleUcW}(Color online). The scheme for fusing three W states of arbitrary size using a Fredkin gate, followed by two fusion gates. A single photon from each state with one ancilla photon are the inputs of the scheme in the spatial modes, $1,2,4$ and $3$, respectively. Note that the single photon from one of the W states is directly input to the fusion gate, whereas one photon from the other two W states is used as a control qubit to swap the ancilla photon with H-polarization with the photon coming from the third W state, if the control photon has V-polarization.}
\end{figure}

In a recent work~\cite{Sahin-NJP-Oct}, it was demonstrated that one can turn the failure case of the fusion gate described above into a success by first acting on the photons coming from the W states by a Fredkin gate (controlled swap gate) whose third input is an ancilla photon in the horizontal (H) polarization, and then feeding the two outputs of the Fredkin gate to the fusion gate. The process can be understood as follows. The failure case of the fusion gate takes place when the photons coming from the W states are both in vertical (V) polarization. If one can switch one of these V-photons with an H-photon then the fusion gate will generate a successful outcome. Thus, using one of these photons as the control qubit of the Fredkin gate, one can swap the ancilla  H-photon with the V-photon coming from the other W state. Thus, the polarization state of the photons input to the fusion gate after the action of the Fredkin gate becomes $\ket{1_H}\ket{1_V}$ which generates a successful outcome at the fusion gate. Consequently, the success probability of fusing W states was increased. Moreover, this scheme enabled starting the W state preparation from Bell states~\cite{Bugu2013A}. However, to reach a W state with a desired size, one has to perform this operation on pairs of W states many times which not only requires a large number of attempts but also a higher communication cost. It was later proposed that using a more complicated scheme involving one Toffoli (CC-NOT) gate, three controlled-not (C-NOT) gates  and two fusion gates operating in parallel, one can fuse four W states at the same time by operating on four photons (one from each of the W states)~\cite{Yesilyurt2013A}. Although efficient, the need for a three-qubit gate and three two-qubit gates  increases the complexity of the setup and makes its experimental realization very difficult.

\begin{figure}[t]
  \centering
      \includegraphics[width=0.45\textwidth]{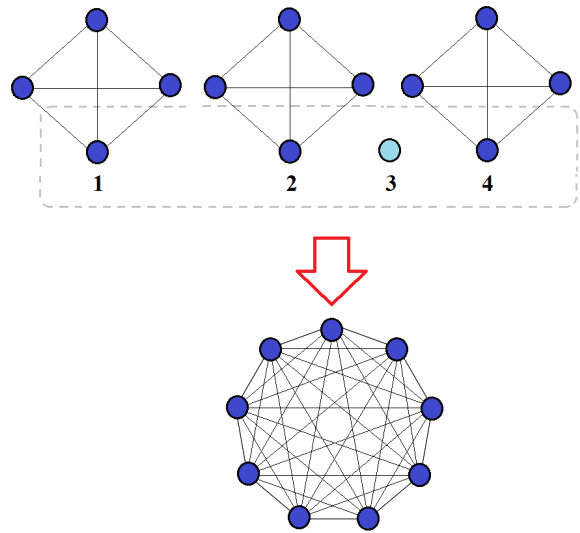}
  \caption{\label{fig:concept}(Color online). Schematic illustration of the proposed fusion scheme. Together with an $|H\rangle$ polarized photon (turquoise), one photon from each $|W_4\rangle$ state (blue) is sent to the fusion scheme shown in Fig.~\ref{fig:AmeleyleUcW} (dashed rectangle). Two photons are destroyed in each fusion gate if the process is successful, creating a $|W_9\rangle$ state. The edges connecting the vertices (qubits) are used to represent the complex entanglement structure of the resources.}
\end{figure}

In this work we provide a scheme to fuse W states but with a much simpler setup compared to the above-mentioned proposals. This new fusion scheme consists of one Fredkin gate and two fusion gates, and uses an ancilla photon in H-polarization, as shown in Fig.~\ref{fig:AmeleyleUcW}. It fuses three W states of arbitrary size with the help of the ancilla state, and requires access to only one qubit of each of the W states. The same scheme can also be used to fuse W-type Bell states to prepare W states with larger numbers of qubits. A schematic illustration of the proposed fusion process is depicted in Fig.~\ref{fig:concept} where three W states of four photons and an ancilla photon are fused to create a W state of nine photons.

\section{Working principle of the \newline proposed fusion scheme}

The fusion scenario we propose is as follows. Alice, Bob and Charlie have W states $|W_n\rangle_A$, $|W_m\rangle_B$ and $|W_t\rangle_C$ of sizes $n,m$ and $t$ respectively, each being greater than or equal to $2$. Their goal is to fuse the states with the help of an ancilla state by sending only one photon from their states to the scheme shown in Fig.~\ref{fig:AmeleyleUcW}.  The states of Alice, Bob, Charlie and the ancilla photon can be written as
\begin{equation}
\!|W_n\rangle_{A} \! =\! {1 \over \sqrt{n} } ( |(n\!-\!1) _H \rangle_a | 1_V \rangle _1\!  +\! \sqrt{n\!-\!1} | W_{n-1}  \rangle_a |1_H \rangle _1 )
\end{equation}
\begin{equation}
\!\!|W_m\rangle_{B} \!=\! {1 \over \sqrt{m} } ( | (m\!-\!1) _H \rangle_b | 1_V \rangle _2 \! + \! \sqrt{m\!\!-\!\!1} | W_{m\!-\!1}  \rangle_b |1_H \rangle _2 )
\end{equation}
\begin{equation}
\!\!|W_t\rangle_{C} \!=\! {1 \over \sqrt{t} } ( | (t\!-\!1) _H \rangle_c | 1_V \rangle _4 \! +\! \sqrt{t-1} | W_{t\!-\!1}  \rangle_c |1_H \rangle _4 )
\end{equation}
and
\begin{equation}
|\Psi\rangle_{anc} \!=\! |1_H\rangle_3.
\end{equation}

According to the scheme shown in Fig.~\ref{fig:AmeleyleUcW}, the photons in spatial modes 1, 2 and 3, from Alice, Bob and the ancilla respectively, are input to the Fredkin gate, while the photon in mode 4 from Charlie is directly sent to fusion gate 2 (FG2). The Fredkin gate swaps the target qubits in modes 2 and 3 if the photon in mode 1 (control qubit) is vertically polarized, {\it i.e.}, if $|i\rangle_1= |1_V\rangle_1$, then $|i\rangle_1|j\rangle_2|k\rangle_3 \rightarrow |i\rangle_1|k\rangle_2|j\rangle_3$, and if $|i\rangle_1= |1_H\rangle_1$, then $|i\rangle_1|j\rangle_2|k\rangle_3 \rightarrow |i\rangle_1|j\rangle_2|k\rangle_3$. After the action of the Fredkin gate the four photons sent to the fusion scheme transform as follows
\begin{eqnarray}
&&|1_H\rangle_{1}|1_{H/V}\rangle_{2}|1_H\rangle_{3}|1_{H/V}\rangle_{4}  \label{N01} \\
&&\qquad \qquad \rightarrow|1_H\rangle_{1'}|1_{H/V}\rangle_{2'}|1_H\rangle_{3'}|1_{H/V}\rangle_{4'} \nonumber
\end{eqnarray}
and
\begin{eqnarray}
&&|1_V\rangle_{1}|1_{H/V}\rangle_{2}|1_H\rangle_{3}|1_{H/V}\rangle_{4} \label{N02} \\
&&\qquad \qquad  \rightarrow|1_V\rangle_{1'}|1_H\rangle_{2'}|1_{H/V}\rangle_{3'}|1_{H/V}\rangle_{4'}. \nonumber
\end{eqnarray}

The photons in modes $1'$ and $2'$ are then sent to fusion gate 1 (FG1) and the photons in modes $3'$ and $4'$ are sent to FG2. Each of the fusion gates has three possible outcomes: Successful (S) when the photons at the inputs of the fusion gate have orthogonal polarization ({\it i.e.} one is H-polarized while the other is V-polarized or vice versa) leading to a coincidence detection as described in Section II; recyclable (R) if both photons are H-polarized; and failure (F) when they are V-polarized. Thus parallel operation of the two fusion gates in the proposed scheme has eight possible outcomes $(S,S)$, $(S,R)$, $(S,F)$, $(R,S)$, $(R,R)$, $(R,F)$, $(F,S)$, $(F,R)$ and $(F,F)$,  which will project the overall state of the remaining $(n+m+t-3)$ photons in spatial modes $a$, $b$ and $c$ to some final state $|\phi\rangle_{abc}$. We will call the operation of the scheme successful when the outcomes of the fusion gates prepare the state $|\phi\rangle_{abc}=|W_{(n+m+t-3)}\rangle_{abc}$.

Note that the outcome $(F,F)$ which requires that all of the four photons at the inputs of the fusion gates are V-polarized never takes place because the ancilla photon is H-polarized. Moreover, the Fredkin gate together with the ancilla H-polarized photon ensures that after the operation of the Fredkin gate, the state at the inputs of FG1 can never be $|1_V\rangle_{1'}|1_V\rangle_{2'}$ and therefore FG1 will never fail completely and the outcomes $(F,S)$, $(F,R)$ and $(F,F)$ never take place. Similarly, FG2 fails completely if its inputs are V-polarized $|1_V\rangle_{3'}|1_V\rangle_{4'}$. This, on the other hand, implies that the ancilla qubit (H-polarized photon) in mode 3 is swapped with the one in mode 2 and hence the control qubit in mode 1 is $|1_V\rangle_{1}$. Thus, in this case, after the operation of the Fredkin gate, the input of FG1 will always be $|V\rangle_{1'}|H\rangle_{2'}$, implying that FG1 will produce a successful output, {\it i.e.} the outcome will be $(S,F)$. Therefore, the cases such as $(F,F)$ and $(R,F)$ can never take place. Thus, the number of possible outcomes in this proposed scheme reduces to five, {\it i.e.} $(S,S)$, $(S,R)$, $(S,F)$, $(R,S)$ and $(R,R)$. Before discussing the final states upon obtaining each of these outcomes at the fusion gates, we note that the outcome $(S,F)$ leads to complete failure of the fusion scheme because in this case the V-polarized photons of the input W states are consumed/destroyed at the detectors of the fusion gates and hence the remaining photons in modes $a$, $b$ and $c$ are all H-polarized, {\it i.e.} $|(n-1)_H\rangle_{a}|(m-1)_H\rangle_{b}|(t-1)_H\rangle_{c}$, and entanglement is completely lost. Thus failure of FG2 leads to complete failure.

It is clear from Eqs.~(\ref{N01}) and (\ref{N02}) that the outcome $(S,S)$ takes place for three out of eight possible inputs to the fusion gates. When the photons in modes 1', 2', 3', and 4' are in the polarization states $|1_H\rangle_{1'}|1_V\rangle_{2'}|1_{H}\rangle_{3'}|1_{V}\rangle_{4'}$, $|1_V\rangle_{1'}|1_H\rangle_{2'}|1_{H}\rangle_{3'}|1_{V}\rangle_{4'}$ or $|1_V\rangle_{1'}|1_H\rangle_{2'}|1_{V}\rangle_{3'}|1_{H}\rangle_{4'}$ the detectors of the fusion gates FG1 and FG2 record a four-fold coincidence detection. These will respectively lead to the output states
\begin{equation}
\frac{\sqrt{n-1}}{\sqrt{nmt}}|W_{n-1}\rangle_{a}|(m-1) _H \rangle_b |(t-1) _H \rangle_c
\end{equation}
\begin{equation}
\frac{\sqrt{m-1}}{\sqrt{nmt}}|(n-1) _H \rangle_a |W_{m-1}\rangle_{b}|(t-1) _H \rangle_c
\end{equation}
and
\begin{equation}
\frac{\sqrt{t-1}}{\sqrt{nmt}}|(n-1) _H \rangle_a |(m-1) _H \rangle_b|W_{t-1}\rangle_{c}.
\end{equation}
Since these are indistinguishable events, the final state $|\phi\rangle_{abc}$ is the superposition of the above expressions. It is straightforward to show that the superposition of the above expressions leads to the W state $|\phi\rangle_{abc}=|W_{n+m+t-3}\rangle_{abc}$ with a success probability of $p_s=(n+m+t-3)/nmt$. Thus for 3 out of the 8 possible inputs to the fusion scheme we are able to fuse the W states together. The other 5 inputs lead to outcomes which we discuss next.

We now see what happens to the final state when both of the fusion gates produce recyclable outputs, {\it i.e.} the outcome  $(R,R)$. In order for this outcome to take place the photons in the inputs of the fusion gates should be all H-polarized. Since the ancilla photon is H-polarized, for this output to take place the photons in modes 1, 2 and 4 should be all H-polarized. With $(R,R)$ as the outcome of the fusion gates, the final state of the remaining photons becomes $|\phi\rangle_{abc}=|W_{n-1}\rangle_{a}|W_{m-1}\rangle_{b}|W_{t-1}\rangle_{c}$ with probability $(n-1)(m-1)(t-1)/nmt$, implying that the states of Alice, Bob and Charlie remain a W state but with one less H-photon. This then allows them to recycle their states back into the fusion scheme once more.

The remaining two cases $(R,S)$ and $(S,R)$ are interesting as they require that one of the fusion gates produce a successful outcome while the other one produces a recyclable outcome.
From the above considerations, it is easy to see that $(R,S)$ will take place if the inputs of the fusion gates are $|1_H\rangle_{1'}|1_H\rangle_{2'}|1_{H}\rangle_{3'}|1_{V}\rangle_{4'}$. In this case, upon detection at the fusion gates, the output state becomes $|\phi\rangle_{abc}=|W_{n-1}\rangle_{a}|W_{m-1}\rangle_{b}|(t-1)_H\rangle_{c}$. Clearly, Charlie's state is completely destroyed because it loses its only V-polarized photon at the detection, while Alice and Bob lose only one H-polarized photon from their state ending up with W states with one less photon. We call this case {\it partial recyclable} (PR) because Alice and Bob can still use their final states but Charlie has to prepare a new one for the next round of the fusion process.

\begin{table}[t]
\centering
\begin{tabular}{l c c c c c c}
\hline
\hline  
Probability                         & Input   & Throughput      & FG1,FG2 & Result  \\ 
\hline \\
 ${(n-1)(m-1)(t-1)  \over nmt  }$ 	& H,H,H,H & \{H,H\},\{H,H\} & R,R & R  \\[1ex]
 ${(n-1)(m-1)       \over nmt  }$ 	& H,H,H,V & \{H,H\},\{H,V\} & R,S & PR \\[1ex]
 ${(n-1)(t-1)       \over nmt  }$ 	& H,V,H,H & \{H,V\},\{H,H\} & S,R & PS \\[1ex]
 ${(n-1)            \over nmt  }$ 	& H,V,H,V & \{H,V\},\{H,V\} & S,S & S1 \\[1ex]
 ${(m-1)(t-1)       \over nmt  }$ 	& V,H,H,H & \{V,H\},\{H,H\} & S,R & PS \\[1ex]
 ${(m-1)            \over nmt  }$ 	& V,H,H,V & \{V,H\},\{H,V\} & S,S & S2 \\[1ex]
 ${(t-1)            \over nmt  }$ 	& V,V,H,H & \{V,H\},\{V,H\} & S,S & S3 \\[1ex]
 ${1                \over nmt  }$ 	& V,V,H,V & \{V,H\},\{V,V\} & S,F & F \\[1ex]
\hline
\hline 
\end{tabular}
\caption{ Truth table of the proposed fusion scheme. ``Input'' indicates the polarizations of the photons at the input of the setup with the associated probabilities, and ``throughput'' indicates the polarizations of photons after the Fredkin gate. Final results are determined according to the joint results of the fusion gates, FG1 and FG2. R stands for the recyclable case; F for the failure case; S1, S2 and S3 for the success cases; PR for the partial recycle case and PS stands for the indistinguishable partial success cases.}\label{T1}
\end{table}

\begin{figure}[b]
  \centering
      \includegraphics[width=0.5\textwidth]{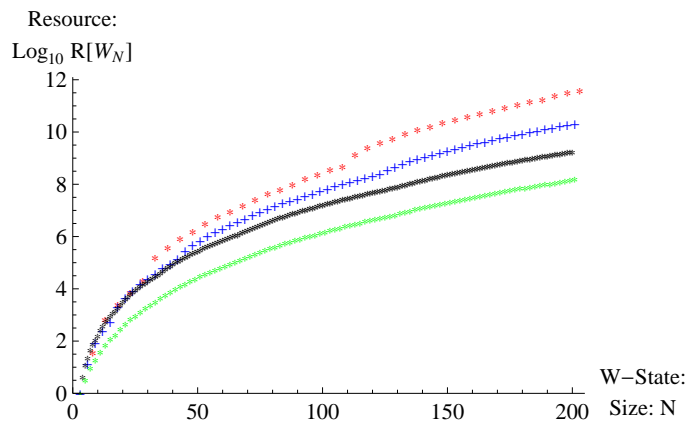}
  \caption{\label{fig:xxx}(Color online). The resource cost of the schemes: Fusing four states (red *) \cite{Yesilyurt2013A}, the present work for fusing three states (blue +); fusing two states (black dots) \cite{Sahin-NJP-Oct}, and fusing two states with enhanced setup (green dots) \cite{Bugu2013A}. Points indicate the actual sizes of the prepared W state.}\label{fig:xxx}
\end{figure}

Finally, the outcome $(S,R)$ can take place for two different states at the inputs of the fusion gates, $|1_H\rangle_{1'}|1_V\rangle_{2'}|1_{H}\rangle_{3'}|1_{H}\rangle_{4'}$ and $|1_V\rangle_{1'}|1_H\rangle_{2'}|1_{H}\rangle_{3'}|1_{H}\rangle_{4'}$. Thus the final state in modes $a$, $b$ and $c$ should be a superposition of the states prepared for these two cases. In the case of $|1_H\rangle_{1'}|1_V\rangle_{2'}|1_{H}\rangle_{3'}|1_{H}\rangle_{4'}$, we see that the photon in mode $1'$ is H-polarized, and therefore no swap has taken place at the Fredkin gate. This implies that the modes before the Fredkin gate are the same as the ones after it, {\it i.e.} $|1_H\rangle_{1}|1_V\rangle_{2}|1_{H}\rangle_{3}|1_{H}\rangle_{4}=|1_H\rangle_{1'}|1_V\rangle_{2'}|1_{H}\rangle_{3'}|1_{H}\rangle_{4'}$. In this case the output state is in the form
\begin{equation}
\frac{\sqrt{(n-1)(t-1)}}{\sqrt{nmt}}|W_{n-1} \rangle_a |(m-1) _H \rangle_b|W_{t-1}\rangle_{c}.\label{N033}
\end{equation}

For the case of $|1_V\rangle_{1'}|1_H\rangle_{2'}|1_{H}\rangle_{3'}|1_{H}\rangle_{4'}$, the control qubit of the Fredkin gate is in the state $|1_V\rangle$ and swapping takes place, however the photons in modes $2'$ and $3'$ are both H-polarized which implies that the photons in modes 2 and 3 were also H-polarized, because this is the only case where these modes can have the same polarization after the Fredkin gate if the control qubit is $|1_V\rangle$. Thus, $|1_V\rangle_{1}|1_H\rangle_{2}|1_{H}\rangle_{3}|1_{H}\rangle_{4}=|1_V\rangle_{1'}|1_H\rangle_{2'}|1_{H}\rangle_{3'}|1_{H}\rangle_{4'}$, implying that the state in modes $a$, $b$ and $c$, upon the outcome at the fusion gates, becomes
\begin{equation}
\frac{\sqrt{(m-1)(t-1)}}{\sqrt{nmt}}|(n-1)_H \rangle_a |W_{m-1} \rangle_b|W_{t-1}\rangle_{c}.\label{N044}
\end{equation}

Since the above two cases are indistinguishable, the output state should be the superposition of the expressions in Eqs.~(\ref{N033}) and (\ref{N044}), which leads to
\begin{eqnarray}
|\phi\rangle_{abc}=|W_{n+m-2} \rangle_{ab}|W_{t-1}\rangle_{c}\label{N055}
\end{eqnarray} with probability $(n+m-2)(t-1)/nmt$. It can be seen that this outcome fuses the states of Alice and Bob to prepare a larger W state. However, Charlie's state is not fused, and it loses one H polarized photon, resulting in a W state which is one photon less than the original state. Therefore, Charlie can recycle his state in a subsequent round of the fusion process. Similarly, now Alice and Bob can use their merged networks in subsequent fusion processes if they want to prepare much larger W states. Therefore, we call this a {\it partial success} (PS) case. Note that although the results of the fusion gates appear as $(R,S)$ in the PR case, {\it i.e.} FG2 actually reports a success, this case is distinguishable from others and it is obvious that the V polarized photon is sent by Charlie. Therefore his state is destroyed. In Table \ref{T1} we give a list of all the possible outcomes and the probabilities of all the cases discussed above.


\section{Cost of preparing larger W states through the proposed fusion scheme}

We now present the results of numerical simulations which compare the resource cost of preparing a W state of a desired size using our proposed fusion scheme and the previously reported schemes, starting with $\ket{W_3}$ states, with and without using the recyclable outcomes. The comparison will be done in particular with the fusion scheme reported in Ref.~\cite{Yesilyurt2013A} which aimed at fusing four W states at once but with more complicated gate combinations, but we also provide a comparison with other setups in the figures. In Ref.~\cite{Sahin-NJP-Oct}, it was shown that in order to prepare a W state of a target size, an optimal way is to fuse states of similar or the same size. Our resource cost analysis is also based on this principle. Using this setup, the optimal way is to fuse three $|W_3\rangle's$ to prepare a $|W_6\rangle$ whereas to fuse two $|W_3\rangle's$ with one $|W_6\rangle$ to prepare a $|W_9\rangle$. As seen in Fig.~\ref{fig:xxx}, even without recycling, the setup presented here (blue + marks) is remarkably more cost efficient than that of Ref.~\cite{Yesilyurt2013A} (red * marks). For the common reachable sizes, $N=33$ and $N=63$, the ratios of the resource costs of present setup to the previous setup Ref.~\cite{Yesilyurt2013A} are $0.232$ and $0.284$, respectively. Note that in this no-recycle strategy, both of the previous setups (black dots for Ref.~\cite{Sahin-NJP-Oct} and green dots for Ref.~\cite{Bugu2013A}) for fusing two states at each step provide a better resource cost efficiency than the present setup. However, below we will see that when the recycling strategy is used, due to the various recyclable cases, the present setup competes with these two previous setups; enlarging the gap between the setup of Ref.~\cite{Yesilyurt2013A}. Even in the no-recycle strategy, to prepare a $N$-qubit W state, we find that a sub-exponential resource cost of $O(\sqrt{N} N^{Log_2N \over k})$  is provided by the setups of  Ref.~\cite{Yesilyurt2013A}, the present work, Ref.~\cite{Sahin-NJP-Oct} and Ref.~\cite{Bugu2013A}, where $k=1.65, 1.9, 2.1$ and $2.45$ respectively.

For the recycling strategy, we follow the approach used in Ref.~\cite{Sahin-NJP-Oct}, where two W states are fused. 
Since it is optimal to fuse states of similar size, in this approach, states are classified into sets which can contain zero, one or two states at a given time. 
The classification is done according to the size of the states, {\it i.e.} a $W$ state of size $N$ belongs to the set $S_l$ where $N \in (2^{l-1}+2, 2^{l}+2] $. 
The first set, $S_0$, can contain only the primary resource states, {\it i.e.} $|W_3\rangle's$. 
Whenever there are two states in a set $S_{l}$, these states are sent to the fusion setup (and $S_{l}$ becomes empty). 
If the result is success, the resultant state belongs to the set $S_{l+1}$. 
If the result is failure, both states are lost. 
If the result is recycle, then the resultant two states may belong to the same set $S_{l}$ or the set $S_{l-1}$. 
Initially all the sets are empty and the resource cost is zero.
The target is to prepare a state which belongs to a set $S_k$ and the final resource cost is the number of $|W_3\rangle's$ that are used during the process. 
The process starts by filling the first set with $|W_3\rangle's$, increasing the cost by one for each. 
We try to fuse these states. 
If they are fused, the resultant state belongs to the second set which now has one state. We then need one more state to try to fuse the states in the second set. 
Therefore we go to the first set and fill it and try to fuse the states there. 
If the result is a success, we have two states in the second set, and we try to fuse them to obtain a state in the third set. We continue this process until we reach the target set $S_k$.
When the recycle case occurs (except for $|W_3\rangle's$ in the first set), the resultant states still belong to some sets, therefore we can still use them without increasing the resource cost. 
The cost is increased by one whenever we need to put a new $|W_3\rangle$ into the first set. 

\begin{figure}[t]
  \centering
      \includegraphics[width=0.5\textwidth]{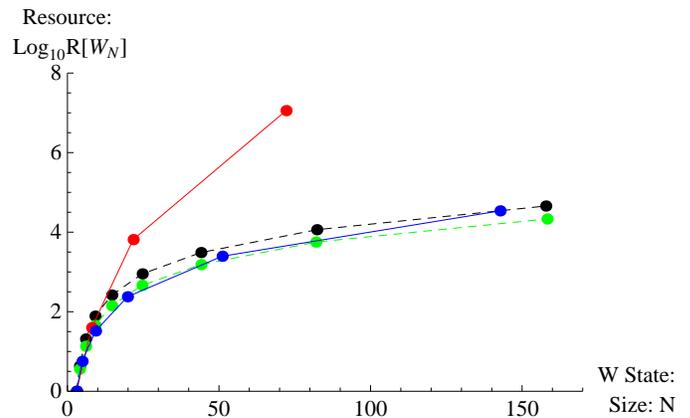}
  \caption{(Color online). The resource costs of the scheme of Ref.~\cite{Yesilyurt2013A} (red solid line), the present work (blue solid line), the scheme of Ref.~\cite{Sahin-NJP-Oct} (black dashed line) and the scheme of Ref.~\cite{Bugu2013A} (green dashed line) versus the sizes of the prepared W states as averaged over 1000 runs.}\label{fig:all-recycled}
\end{figure}

For the numerical results for Ref.~\cite{Bugu2013A}, for the sake of simplicity, we use the same set sizes as in Ref.~\cite{Sahin-NJP-Oct} which does not decrease the cost-efficiency significantly. For the setup of Ref.~\cite{Yesilyurt2013A}, we design five sets to contain zero to four states of possible sizes $N\!=\!3$; $4 \! \leq \! N \! < \! 12$; $12 \! \leq \! N \! < \! 44 $; $44 \! \leq \! N \! < \! 172$; and $172 \! \leq \! N$. For the present setup, we design  six sets of possible sizes $N \! = \! 3$;   $4 \leq  N \! < \! 7$;  $7 \! \leq \! N \! < \! 16$; $16 \! \leq \! N \! < \! 43$; $43 \! \leq \! N \! < \! 124$ and $124 \! \leq \! N$. Adapting the fusion and the recycle processes appropriately, we run each simulation for each set 1000 times and present the results in Fig.4.

It is clear that if recycling is allowed the resource costs of the studied fusion schemes improve significantly, {\it i.e.} the resource costs become lower, compared to the costs when recycling is not allowed. Interestingly, the improvement is the highest for the present scheme that fuses three states simultaneously. With the help of the partial recycle and partial success cases, the present scheme performs as good as the original fusion scheme proposed in Ref.~\cite{Sahin-NJP-Oct} and the enhanced fusion scheme proposed in Ref.~\cite{Bugu2013A}, both of which are restricted to fusing two W-states at a time. Following the exponential fusion strategy that fuses states of the same size,
we find that the schemes presented in Ref.~\cite{Yesilyurt2013A}, here, Ref.~\cite{Bugu2013A} and Ref.~\cite{Sahin-NJP-Oct} respectively prepare 5N-, 3N-, 2N- and N-qubit W-state after N successful fusion steps.


\section{Conclusion}

We have introduced an optical setup for fusing three W states using a scheme which consists of only one three-qubit gate and two basic fusion gates. This is an improvement over previous works which involve fusing four W states with a scheme consisting of a Toffoli gate, two C-NOT gates and two fusion gates. Apart from this reduction in the gate complexity, the present setup is also remarkably cost efficient. From the view point of an experimental implementation of our proposed scheme, although dealing with several W states of many photons is not very practical yet, the fusion of W states of small size can be considered. The main difficulty in implementing the present setup with state-of-the-art photonic technology is the Fredkin gate. However, there are already proposals for the realization of the Fredkin gate using linear optical elements~\cite{Fiurasek2006,Fiurasek2008,Gong2008}. Nonlinear interactions are also promising for such controlled operations in photonics. These interactions require Kerr medium for which realizing an optical quantum Fredkin gate has been proposed~\cite{Milburn1989,Hu2013}. There is also an interesting proposal for constructing an optical Fredkin gate using nanobiophotonics~\cite{Roy2010}. Based on the above-mentioned intense effort focused on realizing the optical Fredkin gate we believe that the experimental realization of our W state fusion scheme is within reach of current technology. Finally, from a more fundamental point-of-view it will be interesting to explore the impact of percolation behavior and quantum critical phenomena in quantum networks~\cite{Acin2007,Kieling2007} using the proposed fusion scheme.

This work was funded by Isik University Scientific Research Fund, Grant Number: BAP-14A101. SKO thanks L. Yang and F. Nori for their continuous support.

\end{document}